\documentclass[sigconf]{acmart}

\AtBeginDocument{%
  \providecommand\BibTeX{{%
    \normalfont B\kern-0.5em{\scshape i\kern-0.25em b}\kern-0.8em\TeX}}}

\setcopyright{acmcopyright}
\copyrightyear{2022}
\acmYear{2022}
\acmDOI{10.1145/3524842.3528435}

\acmConference[MSR '22]{19th International Conference on Mining Software Repositories}{May 23--24, 2022}{Pittsburgh, PA, USA}
\acmBooktitle{19th International Conference on Mining Software Repositories (MSR '22), May 23--24, 2022, Pittsburgh, PA, USA}
\acmPrice{15.00}
\acmISBN{978-1-4503-9303-4/22/05}

\newcommand{\cont}{additional context}
\newcommand{\conts}{additional contexts}
\newcommand\SO{Stack Overflow}
\usepackage{xcolor}

\newcommand\qid[1]{Question\footnote{\url{https://stackoverflow.com/questions/#1}} #1}
\newcommand\tc[1]{\raisebox{.5pt}{\textcircled{\raisebox{-.9pt} {#1}}}}

\usepackage{multirow}
\usepackage{booktabs}
\usepackage{tablefootnote}
\usepackage{graphicx}

\usepackage[absolute,overlay]{textpos}
\usepackage{tikz}
\usepackage{xcolor}
\usepackage{calc}

\newcounter{findingctr}

\newcommand{\finding}[2]{\refstepcounter{findingctr}\emph{#1:\label{box:#2}}}

\newlength{\boxw}
\newlength{\boxh}
\newlength{\shadowsize}
\newlength{\boxroundness}
\newlength{\tmpa}
\newsavebox{\shadowblockbox}

\newenvironment{findingenv}[2]%
{\vspace{0.2cm}\noindent
\begin{lrbox}{
\shadowblockbox
}
\begin{minipage}{.98\columnwidth}
\finding{#1}{#2}~}%
{\end{minipage}\end{lrbox}%
\settowidth{\boxw}{\usebox{\shadowblockbox}}%
\settodepth{\tmpa}{\usebox{\shadowblockbox}}%
\settoheight{\boxh}{\usebox{\shadowblockbox}}%
\addtolength{\boxh}{\tmpa}%
\begin{tikzpicture}
\addtolength{\boxw}{\boxroundness * 2}
\addtolength{\boxh}{\boxroundness * 2}

\foreach \x in {0,.05,...,1}
{
\setlength{\tmpa}{\shadowsize * \real{\x}}
\fill[xshift=\shadowsize - 1pt,yshift=-\shadowsize + 
1pt,black,opacity=.04,rounded corners=\boxroundness] 
(\tmpa, \tmpa) rectangle +(\boxw - \tmpa - \tmpa, \boxh - \tmpa - 
\tmpa);
}

\filldraw[fill=white!50, draw=black!80, rounded corners=\boxroundness] (0, 
0) rectangle (\boxw, \boxh);
\draw node[xshift=\boxroundness,yshift=\boxroundness,inner sep=0pt,outer 
sep=0pt,anchor=south west] (0,0) {\usebox{\shadowblockbox}};
\end{tikzpicture}\vspace{0cm}%
}

\begin{document}

\title{Does This Apply to Me? An Empirical Study of Technical Context in Stack Overflow}

\author{Akalanka Galappaththi}
\affiliation{%
  \institution{University of Alberta}
  \city{Edmonton}
  \state{Alberta}
  \country{Canada}
  \postcode{T6G 2E8}}
\email{akalanka@ualberta.ca}

\author{Sarah Nadi}
\affiliation{%
  \institution{University of Alberta}
  \streetaddress{2-41 Athabasca Hall}
  \city{Edmonton}
  \state{Alberta}
  \country{Canada}
  \postcode{T6G 2E8}}
\email{nadi@ualberta.ca}

\author{Christoph Treude}
\affiliation{%
  \institution{University of Melbourne}
  \streetaddress{Grattan Street, Parkville}
  \city{Victoria}
  \country{Australia}}
\email{christoph.treude@unimelb.edu.au}

\renewcommand{\shortauthors}{Galappaththi et al.}

\begin{abstract}
Stack Overflow has become an essential technical resource for developers.
However, given the vast amount of knowledge available on Stack Overflow, finding the right information that is relevant for a given task is still challenging, especially when a developer is looking for a solution that applies to their specific requirements or technology stack. 
Clearly marking answers with their \textit{technical context}, i.e., the information that characterizes the technologies and assumptions needed for this answer, is potentially one way to improve navigation.
However, there is no information about how often such context is mentioned, and what kind of information it might offer.
In this paper, we conduct an empirical study to understand the occurrence of technical context in Stack Overflow answers and comments, using tags as a proxy for technical context. 
We specifically focus on \textit{additional context}, where answers/comments mention information that is not already discussed in the question.
Our results show that nearly half of our studied threads contain at least one additional context. We find that almost 50\% of the additional context are either a library/framework, a programming language, a tool/application, an API, or a database.
Overall, our findings show the promise of using additional context as navigational cues.

\end{abstract}




\maketitle
\section{Introduction}

\begin{figure}[t!]
 \centering
 \includegraphics[scale=0.6]{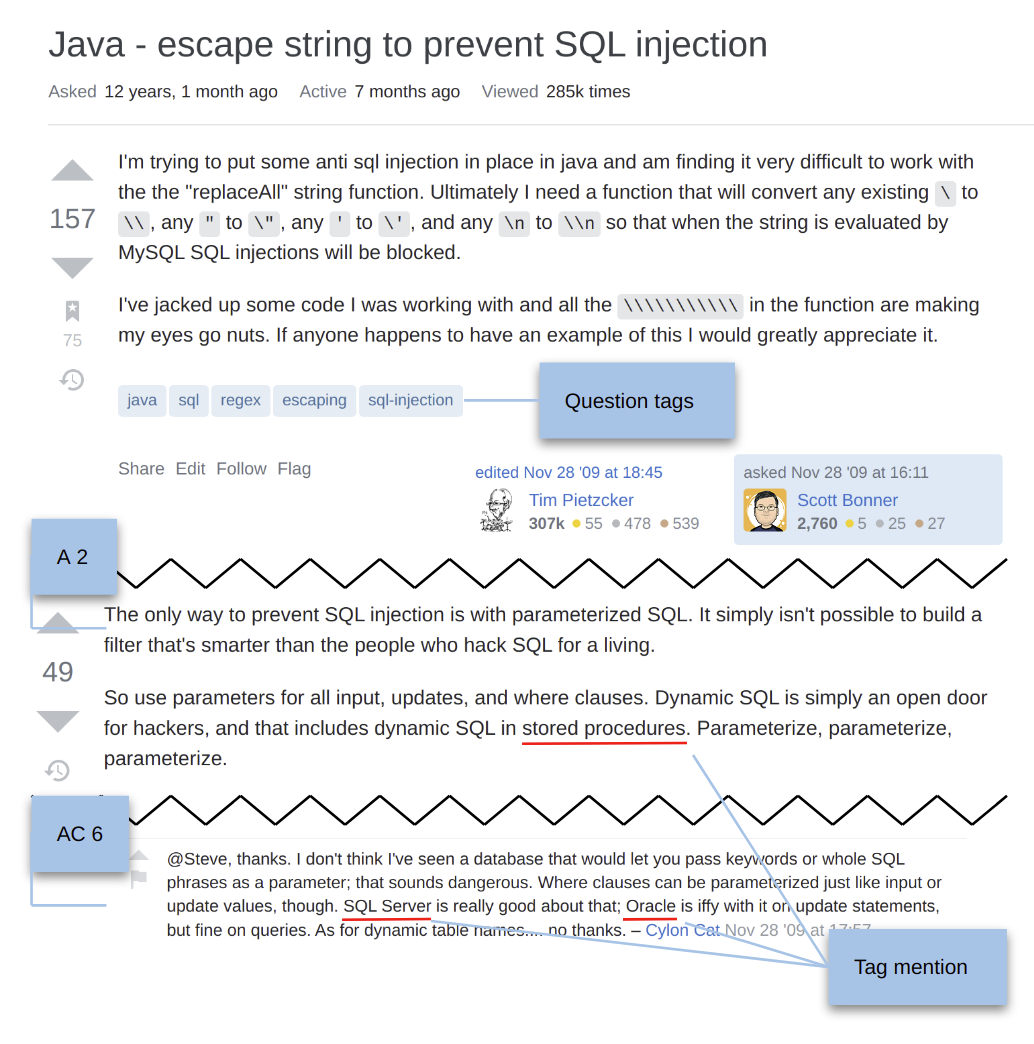}
 \caption{Motivating example based on \SO{} Question 1812891. Underlined words are \SO{} tags that appear as plain text in sentences.}
 \label{fig:motivating_example}
\end{figure}

Stack Overflow is a popular Q\&A website among software developers, with about 22 million questions and 32 million answers. When performing their tasks, developers typically resort to \SO{} to quickly find relevant information~\cite{SODisAndroid:2014, SOCodeReuse:2017}, especially given the pressure of quickly releasing software~\cite{SWMag:2008}. 

Given the vast amount of knowledge on Stack Overflow, finding the right piece of information that is relevant for a developer's task could be challenging though, especially in long threads\footnote{We use the term thread to refer to a \SO{} question and all its answers and comments.}~\cite{Xu:2017}.
Squire and Funkhouser found that 95\% of the \SO{} answers consist of 75\% text~\cite{SOCodetoText:2014} while Nadi and Treude found that 37\% of all questions on Stack Overflow have more than one answer, with 789 characters as the average length of an answer~\cite{Nadi:2020}.
Thus, since most threads contain large amounts of text and explanations, a developer could miss important information that is relevant to what they are looking for.
This is especially true given the current limited visual indicators that may help developers focus their attention on relevant information. 
Current visual indicators on Stack Overflow include (1) question tags that typically indicate the technologies related to a question, (2) question and answer scores that show how the community evaluate a given question or answer, (3) the accepted answer checkmark which indicates which answer solved the original poster's problem, and (4) the question/answer activity indicator which shows the change history of the post \cite{SOTour}. None of these visual cues provide quick indications of the content of a given answer or comment to help users navigate the provided information.

Figure~\ref{fig:motivating_example} shows an example \SO{} thread.
This question is tagged with the \emph{java}, \emph{sql}, \emph{regex}, \emph{escaping}, and \emph{sql-injection} tags.
These question tags typically indicate the technologies related to the question of the original poster and also help with the search functionality on \SO{}~\cite{SOGuide}.
By looking at the question tags in Figure \ref{fig:motivating_example}, users can clearly understand the technologies used in the question without even reading the question body.
In other words, these question tags inform users about the context of a question.
While context is an overloaded term in computer science~\cite{Context:2001}, in this paper, we define \textit{technical context} (or \textit{context} for short throughout the paper) as information that characterizes the technologies and assumptions of a given question or answer.
In this example thread, the question is relevant only to developers who are using Java, and not, for example, any other programming language.
While questions present their context information with clear visual cues, other elements of a thread do not have such visual indicators of their contexts. For long threads that have multiple answers, answers with long textual descriptions, or answers with long code snippets, it would be useful for developers to quickly identify the answer context. By knowing the answer context, a developer could make an informed decision on whether to read or skip the answer.

For example, the thread in Figure \ref{fig:motivating_example} has fourteen answers. Answer 2 (A 2) mentions context information such as \emph{SQL injection}, \emph{SQL}, and \emph{stored procedures} while Comment 6 on Answer 2 (AC 6) contains \emph{SQL}, \emph{SQL server}, and \emph{Oracle}. 
Since \emph{SQL} and \emph{SQL injection} are already part of the question context, a user reading this thread already expects that answers will discuss or mention this context.
On the other hand, \emph{stored procedures}, \emph{SQL server} and \emph{Oracle} are different contexts from those in the question, which we refer to as \emph{\cont}. 
Users looking for solutions for these specific techniques or technologies may want to quickly jump to the relevant answers that discuss them.
Unfortunately, with such \cont{} information hidden within plain-text sentences, users will not be able to navigate quickly to or away from these answers or comments.
It is worth noting that the accepted answer of the thread in Figure~\ref{fig:motivating_example} does not have any \conts{}. However, users may focus their attention on it because of the visual indicator (i.e. checkmark) provided by \SO, and miss relevant \conts{} in other answers.

Overall, identifying the \cont{} of answers and comments and clearly marking these with visual cues (e.g., answer tags) could be a potential avenue for improving the navigation of \SO{} threads.
Similar to a user navigating to an accepted answer as it has a checkmark, an answer tag will help a user to navigate to an answer with a technical context that is related to their particular context.
However, we first need to understand whether such additional contextual information is present in \SO{} answers and comments and what the nature of such information is.
Thus, in this paper, we conduct an empirical study of additional context in \SO{} answers and comments to answer the following research questions:

\begin{itemize}
	\item[RQ1:] \textbf{How frequently is \cont{} available in \SO{} answers and comments?} If answers and comments never have additional context, then the whole idea of extracting and emphasizing this information is not feasible. On the other hand, if there is too much additional context, then this may result in too many visual cues that make a thread cluttered. Thus, it is important to first understand how often additional context occurs and where it occurs in a thread.
	\item[RQ2:] \textbf{What types of \cont{} are available in \SO{} threads?} In general, there may be various types of context that exist in a thread (e.g., API, programming language, operating system etc.). It is important to understand what types of \cont{} appear in threads to decide what may be useful for navigation.
	\item[RQ3:] \textbf{What is the purpose of \cont{} information?} While understanding the general type of \cont{} is important, we also want to understand why a Stack Overflow answer or comment mentioned the additional context (i.e., what purpose does this \cont{} serve in this particular thread). 

\end{itemize}

To answer these research questions, we perform a quantitative and qualitative empirical study of 207 threads from three technologies on \SO, \emph{Json}, \emph{Django}, and \emph{Regex}. Our sample contains 488 answers, 468 answer comments, and 550 question comments, and a total of 3,504 sentences.
To scope our study, we use the technologies in existing \SO{} tags as a proxy for contextual information and automatically search for the occurrence of these tags in the answers and comments of our studied threads.
We then manually verify if the automatically detected tag occurrences are properly used in the marked sentences and whether these tag mentions represent additional contextual information (i.e., do not overlap with the question context).
We then use open-coding card sorting to identify categories of \cont{} as well as their purpose in the thread. 

Our results show that approximately half of the studied threads contain at least one sentence with additional context. From those, 50\% of the additional contexts belong to categories such as library/framework, programming language, tool/application, API, and database. 
We also find that 20\% of sentences with additional context provide direct solutions or solution conditions that would be valuable for a developer navigating this thread.
We provide a discussion of the implications of our results for augmenting \SO{} threads with navigational cues based on \conts{}.

\section{Related Work}

While, to the best of our knowledge, there is no existing research that specifically focuses on finding \conts{} on \SO, there are several studies and related research that we build our motivation and initial insights on. We review existing work on identifying relevant information and navigating to relevant information on \SO.

\paragraph{Identifying relevant information on \SO:} Nasehi et al. conducted a qualitative study to find good code examples on \SO{} \cite{SOGoodCode:2012}.
To find characteristics of good code examples, the authors studied the code snippets, their surrounding text, as well as answer comments. They found that answers often offer alternative solutions and discuss various API performance issues and that comments provide further explanations. These findings motivate us to further study the textual content in \SO{} threads in order to identify \conts{} in sentences, with the eventual goal of helping users navigate this information. 

Treude and Robillard argued that \SO{} users often skip the text surrounding code snippets when they cannot clearly determine whether it explains the code in general or answers the specific requirements in the original question \cite{SOCodeFrag:2017}. Even though the authors did not focus on the surrounding text in their study, this argument supports our assumption of the availability of different contextual information. Imtiaz et al. explore \SO{} questions related to static analysis tools to understand the challenges developers face when responding to these tools' alerts \cite{SAT:2019}. While categorizing the questions they studied, the authors found that some of the questions contain different contextual information. For example, direct solutions and conceptual knowledge related to the problem were present in a single thread.  Therefore, emphasizing contextual information in threads could help users quickly determine the context and decide whether to skip or read the descriptions.

Treude and Robillard found that sentences extracted from \SO{} threads could complement API documentation \cite{SOAPIAug:2016}. Baltes et al. studies links to documentation in \SO{} threads \cite{SODocRef:2020}. Those two studies have fundamentally different goals, one to enrich software documentation with missing information and the other to improve information diffusion. However, some of the sentences that they found useful in both studies contain technical context that motivates us to identify contextual information in such sentences that could help users to navigate through threads. 

\paragraph{Navigating to relevant information on \SO:} Due to the information overload, navigating to relevant information on \SO{} threads could be challenging. A recent survey by Chattarjee et al. shows that too much text in answer descriptions is one of the reasons that slows down developers from identifying relevant solutions on \SO{} \cite{SOFixErrors:2020}. As an interpretation of their results, the authors suggest that a solution is to highlight the relevant content for users. Our work has a similar motivation but we specifically focus on \conts{} in this information. Along similar lines, Xu et al. also argue that finding relevant information in a long post is difficult, but instead of highlighting relevant information, they generate a summary from the answers \cite{Xu:2017}.

Nadi and Treude extracted essential sentences that help users navigate \SO{}  threads \cite{Nadi:2020}. They define \textit{essential sentences} as those that allow users to determine whether an answer should be read or skipped. 
The authors compare four automated techniques for identifying essential sentences, using ratings from survey participants as the ground truth.
This is the closest work to ours: we share the end goal of identifying content that could potentially help users to quickly navigate through long threads; additionally, some of the \emph{essential sentences} they found contain \conts, although the authors do not recognize it as such. Their results show that some of the highly rated sentences by survey participants were missed by all four techniques, while others were captured by only one out of the four techniques. 
The authors concluded that there is no single superior technique for capturing essential sentences for navigation.
The findings of their work indicate that before automatically capturing navigational cues, we need to better understand what information can potentially be used as navigational cues. Therefore, in this paper, we take a step back and focus on one category of well-defined information, technical context, analyze its presence in \SO{} threads, and its potential to be used as navigational cues.

\section{Methodology}

\begin{figure}[t!]
 \centering
 \includegraphics[scale=0.35]{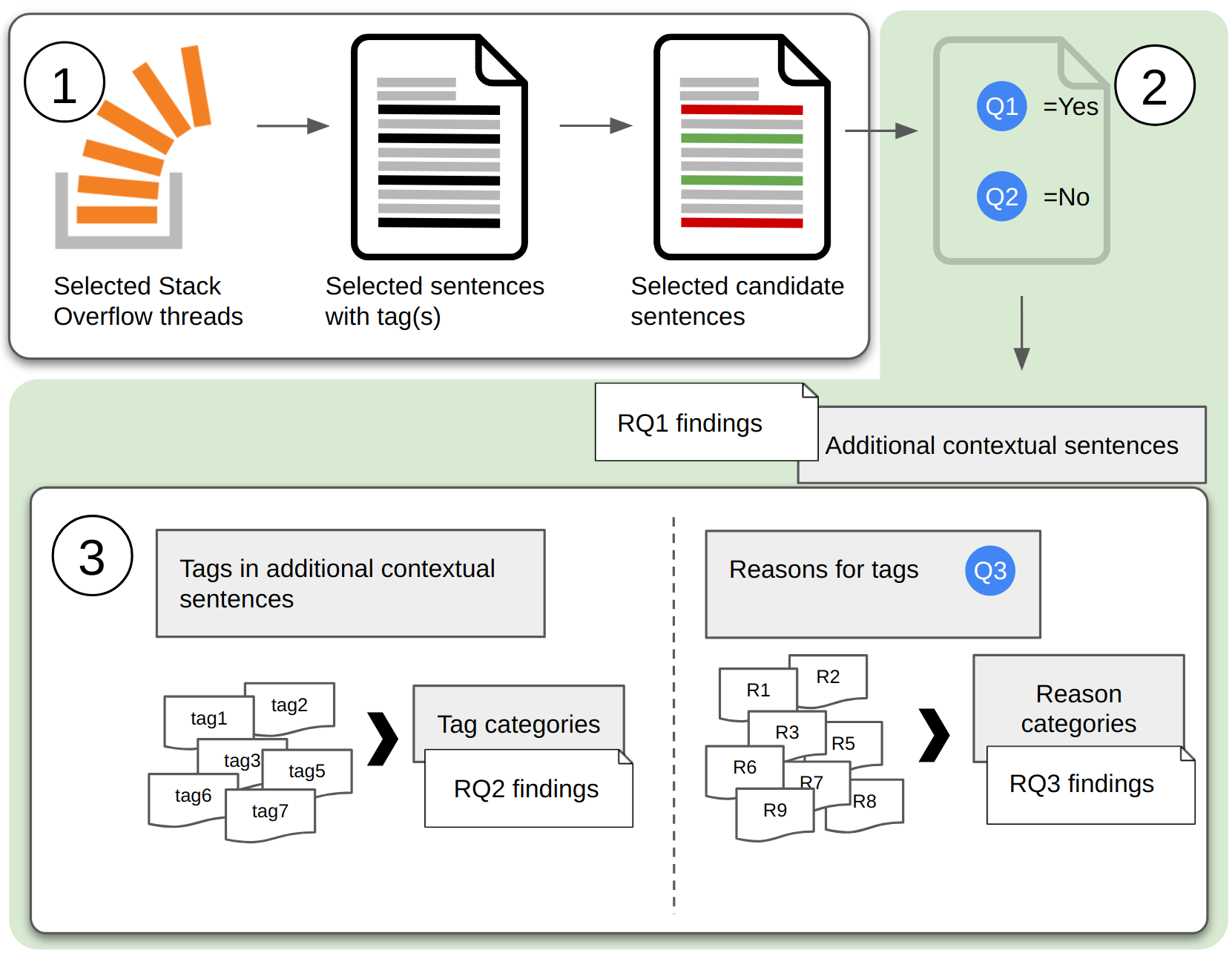}
 \caption{Overview of the research methodology.}
 \label{fig:overview2}
\end{figure}

Figure \ref{fig:overview2} illustrates the three phases of collecting the required data and information for our empirical study. In Phase \tc{1}, we use automated criteria to analyze the selected \SO{} threads and identify sentences that potentially contain additional context. 

In Phase \tc{2}, we manually review the candidate sentences from Phase \tc{1} and confirm those that contain additional context, which allows us to answer RQ1.

In Phase \tc{3}, we perform open card sorting to categorize the types of tags that appear as additional context as well as the reasons for mentioning these additional contexts, which allows us to answer RQ2 and RQ3. 

\subsection{Phase \tc{1} Automatically select sentences for manual review}

The goal of Phase \tc{1} is to automatically identify sentences that potentially contain additional context. 
We will then manually review these candidate sentences in Phase \tc{2}.
Phase \tc{1} has three steps. First, we select a sample of \SO{} threads for our empirical study. Second, we automatically identify sentences in answers and comments that contain technical context.
We define \textit{technical context} as information that characterizes the technologies and assumptions of a given question or answer. We use \SO{} tags as a proxy for technical context.
Third, we automatically compare the technical context identified in the answer/comment sentences to the set of tags that exist in the question; this allows us to filter out sentences whose context already overlaps with that of the question.
The output of Phase \tc{1} is a set of candidate sentences with potential additional context that we manually review in Phase \tc{2}.
 
\subsubsection{Select threads for empirical study}

To investigate the availability of additional contexts, we will need to eventually manually examine the sentences in \SO{} answers and comments. Manual examination of \SO{} threads is time-consuming and requires domain expertise. Since \SO{} has over 21 million questions, to balance manual effort and representativeness, we limit our study to questions that belong to one of three tags: \emph{json}, \emph{regex}, and \emph{django}. 
We selected these tags to diversify the types of threads we study. \emph{Json} is a general data exchange format supported by many programming languages. \emph{Regex} is also supported by many programming languages for pattern matching in strings. \emph{Django} is a popular open-source Python web framework. 
We use the \SO{} dump\footnote{https://archive.org/details/stackexchange} from September 7, 2021, to collect all question IDs that belong to these three tags. 
The second column of Table~\ref{tab:so_q_stat} shows the total number of questions available in the \SO{} dump for our three selected tags.

From the pool of available question IDs, we remove questions that do not have any answers. Note that a question could be tagged with more than one of our selected three tags (e.g., \SO{} \qid{8586346} has both \emph{django} and \emph{regex} as tags). In such cases, we remove duplicates by attaching this question ID to the set of questions related to only one of the tags.
The third column of Table~\ref{tab:so_q_stat} shows the corresponding number of considered questions after filtering and removing duplicates.
Given the large number of threads available, we randomly select a representative sample of question IDs for our study. When calculating the sample size, we choose 90\% confidence interval and 10\% margin of error, so that our sample size is practical for manual examination. The resulting sample size is 69 threads from each tag, for a total of 207 threads. These 207 threads contain 488 answers with 1,644 sentences, 468 answer comments with 1,064 sentences, and 550 question comments with 776 sentences.

\begin{table}[t!]
    \centering
    \caption{Statistics of number of questions}
	\begin{tabular}{lrr}
		\toprule
		 \multicolumn{1}{l}{\textbf{Tag}} & 
		 \multicolumn{1}{r}{\textbf{Questions}} &
		 \multicolumn{1}{r}{\textbf{Questions w/answers}} \\
		 & & \multicolumn{1}{r}{\textbf{\& no duplicates}} \\
		\midrule
		Json & 264,898 & 226,287 \\
		Django & 212,850 & 168,384 \\
		Regex & 168,969 & 158,490 \\
		\bottomrule	
	\end{tabular}
	\label{tab:so_q_stat}
	
\end{table}

\subsubsection{Identify sentences with technical context}

To reduce the burden of manually reading through each sentence in our sample to determine if it contains contextual information, we design an automated approach to identify candidate sentences for review.

We use \SO{} tags as a proxy for technical context.
However, using tags to capture contexts poses two problems. The first problem is that \SO{} tags are user created content. Therefore, there is no guarantee that a tag has a description or that the tag description has enough information to determine what kind of technology it represents.  
If a tag does not have a description it is not possible for us to determine whether that tag is used as intended in a sentence.  
The second problem is that \SO{} currently has over sixty thousand tags. While some of these tags represent technical contexts, the same tag can be used as a regular word in a sentence. \emph{Flow}, \emph{monitor}, and \emph{back} are a few examples that fall under this category. 

To solve the above problems, we leverage the \emph{Witt} taxonomy of \SO{} tags by Nassif et al.~\cite{Witt:2019}. 
The authors automatically analyzed tag descriptions, tag wikis, and Wikipedia pages of a tag to identify a tag's category~\cite{Witt:2019}. Using this information, they then created a taxonomy of \SO{} tags. Table \ref{tab:witt_ex} shows examples of tags and corresponding categories extracted from the \emph{Witt} taxonomy. However, given the automated process, some of the categories in the taxonomy are not meaningful. For example, ``project'', ``software'', ``developed'', ``framework'', ``deprecated'', and ``abandoned'' are the categories for \emph{moonlight} in \emph{Witt}\footnote{https://witt.herokuapp.com/}. According to the \SO{} tag description, \emph{moonlight} is the Linux port for Microsoft's cross-browswer plugin called Silverlight. Considering this tag description, ``project'', ``software'', and ``framework'' are more suitable as categories while ``developed'', ``deprecated'', and ``abandoned'' are more suitable as attributes. %
After investigating the taxonomy, we observe that some tag categories such as \emph{programming-language}, \emph{framework}, and \emph{library} have more than 200 corresponding tags, while other categories have very few tags. In general, we find that the category frequency distribution is right skewed with the median of 2 tags per category and that low count categories in the long tail of the distribution were not meaningful (e.g., \textit{abandoned} and \textit{end-user}). To ensure that we consider only meaningful categories, we discard categories that describe only a limited number of tags; this amounted to discarding 99\% of the categories. The remaining 1\% contains 29 categories in which each category has more than 230 tags. This subset of Witt categories contains 14,536 tags, representing $\sim$35\% of the total number of tags in the taxonomy.
We find that our selected subset of tags does not capture tags like \emph{push}, \emph{drop}, and \emph{background} as technical contexts in sentences, but is still able to capture tags that represent technologies such as \emph{java}, \emph{node}, and \emph{oracle}.

\begin{table}[t!]
    \centering
    \caption{Selected tags and their categories extracted from \emph{Witt} taxonomy~\cite{Witt:2019}}
	\begin{tabular}{llll}
		\toprule
		\textbf{Tag} & \textbf{Category} & \textbf{Tag} & \textbf{Category} \\
		\midrule
		django-1.10 & web-framework & jackson & api  \\
		json & format & r-tree & data-structure \\
		linuxmint & os & oracle & database  \\
		gson & library & notepad++ & editor\\
		\bottomrule	
	\end{tabular}
	\label{tab:witt_ex}
	
\end{table}

\SO{} tags can contain one word (e.g., \emph{java}) or multiple words (e.g., \textit{sql-server}). 
Thus, to capture tags mentioned in sentences, we have to consider different word combinations. We generate unigrams, bigrams, and trigrams from the words in sentences to compare with the filtered tag list. To demonstrate this, Table \ref{tab:tag_cap} contains two sentences from AC 6 of our motivating example in Figure~\ref{fig:motivating_example}. Generally, multiword \SO{} tags use hyphens to separate each word (e.g. \emph{java-collection-api}, \emph{python-3.6}). However, some tags do not follow the same format (e.g \emph{android4.0.3}). To capture tags with different formats, we first replace all hyphens with spaces. Then, we use the following regular expression \verb|([a-z]+)(\d+(\.\d+)*)([a-z]+)?| to capture version numbers that appear after text or in the middle of text. For example, occurrences of \emph{java 7} or \emph{java-7} or \emph{java7} will all be treated as \emph{java 7} during the matching process. If any unigram, bigram, or trigram matches a tag, we capture it as a technical context. The underlined unigram and bigram in Table \ref{tab:tag_cap} are the tags mentioned in the two sentences from AC 6 in Figure~\ref{fig:motivating_example}. 

\begin{table}[t!]
    \centering
    \caption{Capturing tags mentioned in sentences}
	\begin{tabular}{lp{6cm}}
		\toprule
		\textbf{Sentence} & SQL Server is really good about that\\
		\midrule
		\textbf{unigram} & sql, server, is, really,... \\
		\textbf{bigram} & \underline{sql-server}, server-is, is-really, ...  \\
		\textbf{trigram} & sql-server-is, server-is-really, ...  \\
		
		\toprule
		\textbf{Sentence} & Oracle is iffy with it on update ...\\
		\midrule
		\textbf{unigram} & \underline{oracle}, is, iffy, ... \\
		\textbf{bigram} & oracle-is, is-iffy, ...  \\
		\textbf{trigram} & oracle-is-iffy, is-iffy-with, ...  \\
		\bottomrule	
	\end{tabular}
	\label{tab:tag_cap}
	
\end{table}

\begin{figure}[t!]
 \centering
 \includegraphics[scale=0.6]{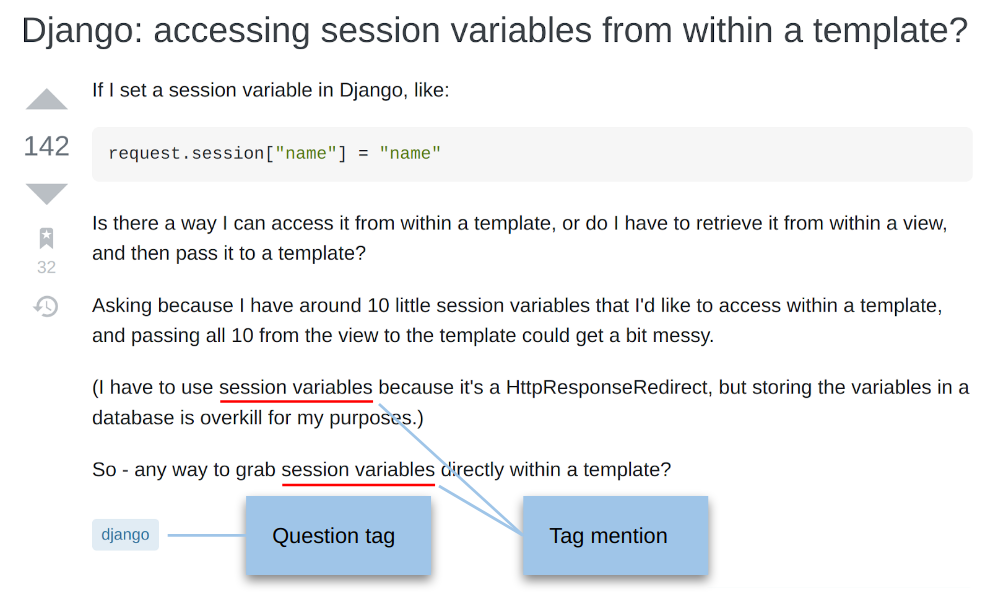}
 \caption{\SO{} 2551933 has only one question tag, \emph{Django}. However, the question body contains \emph{session variables} which is a \SO{} tag.}
 \label{fig:q_context}
\end{figure}

\paragraph{Identify sentences with additional context}
Recall from the introduction that we are interested in identifying sentences with \cont, that is contextual information that is not already included in the question.
When posting questions, \SO{} users can include tags to indicate the context of the question.  Our running example (Figure \ref{fig:motivating_example}) has five tags, \emph{java}, \textit{sql}, \textit{regex}, \textit{escaping}, \textit{sql-injection}.
Thus, identifying that an answer or comment sentence mentions java, for example, is not particularly useful for navigation since it is natural that the whole thread is about java.
However, there may also be technical contextual information mentioned in the question text without a corresponding question tag.
For example, Figure~\ref{fig:q_context} shows \qid{2551933} tagged with \emph{django}. The question body also contains \emph{session-variables}, which is a \SO{} tag and thus would be considered technical context based on our proxy.
Thus, similar to how the question tag \emph{django} is expected to be mentioned in many places in this thread, it is also likely that many answers and comments will mention \emph{session-variables}. The red underlines in Figure \ref{fig:q_context} shows the tag mentioned as plain text in the question body.

Thus, to identify sentences with additional context, we first compile the question's technical context by combining any tags mentioned in the question body as well as the explicitly included question tags.
We use the same subset of tags extracted from \emph{Witt} to capture tags mentioned in the question body.
Given the identified answer and comment sentences with technical context, we filter out any sentences where the identified context overlaps with the question's context.

At the end of Phase \tc{1}, we are left with a set of automatically identified sentences that likely contain additional context. However, we still need to manually review these sentences for confirmation.

\subsection{Phase \tc{2} Manual Confirmation of Additional Context}
\label{sec:phase2}

While Phase \tc{1} automatically identifies candidate sentences that match our definition of additional context, the automated detection process can flag sentences that do not actually provide additional context. 

For example, the Stack Overflow tag \texttt{super} refers to the programming language keyword used to invoke members of a super class.
Our automated detection process would flag a sentence such as \textit{In the super worst case, ...}. However, this sentence does not actually contain any technical context.
Thus, we need to first manually analyze each flagged sentence to make sure that the tag is used as intended and that this sentence actually contains additional context.
Additionally, in this manual review phase, we perform a qualitative analysis on the reason why the additional context is mentioned.
Specifically, we answer the following questions during the manual review phase:

\begin{itemize}
	\item[(Q1)] Is the tag used as intended in the tag description? \textit{Yes/No} 
	\item[(Q2)] Does this tag overlap with the immediate parent's context? \textit{Yes/No} 
	\item[(Q3)] Why is this ``tag/context'' mentioned in this thread? \textit{Free form text} 
\end{itemize}

We use Q1 and Q2 to confirm sentences with additional context, which we then use to answer RQ1 as shown in Figure~\ref{fig:overview2}.
We first explain the coding guide for answering these questions and then describe the process we used to distribute sentences across multiple authors for review. In Phase \tc{3}, we perform open card sorting \cite{OpenCoding:2016} of the reviewers' answers of Q3 (collected in Phase \tc{2}) to determine categories of reasons. We then ask the reviewers to go through the sentences again and use these fixed categories to answer Q3. 

\begin{table*}[t]
    \centering
    \caption{Examples of annotation process}
    \label{tab:annotation_ex}
	\begin{tabular}{lp{8.9cm}cccc}
		\toprule
		\textbf{\#} & \textbf{Sentence} & \textbf{Tag} & \textbf{Used as} & \textbf{Overlap (Q2)} & \textbf{Additional} \\
		& & & \textbf{intended (Q1)} & & \textbf{context}\tablefootnote{Answer of this column determined by the outcome of Q1 and Q2. It is shown here only for the demonstration, but never used during the annotation.}\\
		\midrule
		
		1 &in Django 1.10 , the \verb|django.template.context_processors.request| was already in the setting file : D  & django 1.10 & Yes & No & Yes\\
		
		2 & Ultimately, use the querystring method to get around the dot in the route  & dot & No & - & - \\
		
		3 & Because you are deploying on Heroku, you need to use something like white noise to serve your static files. & static & Yes & Yes & No \\
		
		\bottomrule	
	\end{tabular}
\end{table*}

\paragraph{Coding guide for annotating question 1 (Q1):} We use the \SO{} tag description to determine whether a tag is used as intended in a sentence. We create a guideline to help annotators when they face ambiguous cases. If a tag has no description, we answer \textit{No} to Q1. For example, \emph{defaults} and \emph{data-lake} have no description. When those tags appear in sentences, we select \textit{No} to avoid subjective interpretation of the tag's intention. If the tag is part of a URL (e.g.\emph{php} in \verb|owasp.org/index.php/|), part of a file name (e.g. \emph{jackson} in \verb|jackson-module-jaxb-annotations-2.3.2|), or a path (e.g. \emph{html} in \verb|/var/www/html/sandbox/sandbox/wsgi.py|), we choose \textit{No} to Q1 as well.
Table \ref{tab:annotation_ex} shows three sentences extracted from the sample of threads that we annotated for this study. We only show the responses for the first two questions (Q1 and Q2) due to space limitation and clarity. 
The tag column shows the tag in the sentence as automatically captured by Phase \tc{1}.
For example, the captured tag is \emph{django 1.10} in sentence 1. Since this sentence discusses the settings file content in a particular Django version (i.e. 1.10), an annotator would answer `Yes' to whether the tag is used as intended.
In sentence 2, the detected \emph{dot} tag in the sentence does not match with the tag description. Therefore, the answer to Q1 would be \textit{No}. Note that if a reviewer answers \textit{No} to Q1, they do not need to answer the rest of the questions, since the rest of the questions only apply if the tag is used as intended.

\paragraph{Coding guide for annotating question 2 (Q2):} When answering Q2, we consider the content hierarchy of a \SO{} thread. Our automated process in Phase \tc{1} already does a certain amount of filtering. For example, if a tag mentioned in a sentence is in the question context, then that sentence is filtered out. For simplicity and precision, we did not automatically process the code snippets or images in the question, because we focused on textual content.
However, since there may be cases where a technical context is mentioned in a code snippet in the question,
we need to manually verify if there is an overlap. 

\begin{figure}[t!]
 \centering
 \includegraphics[scale=0.4]{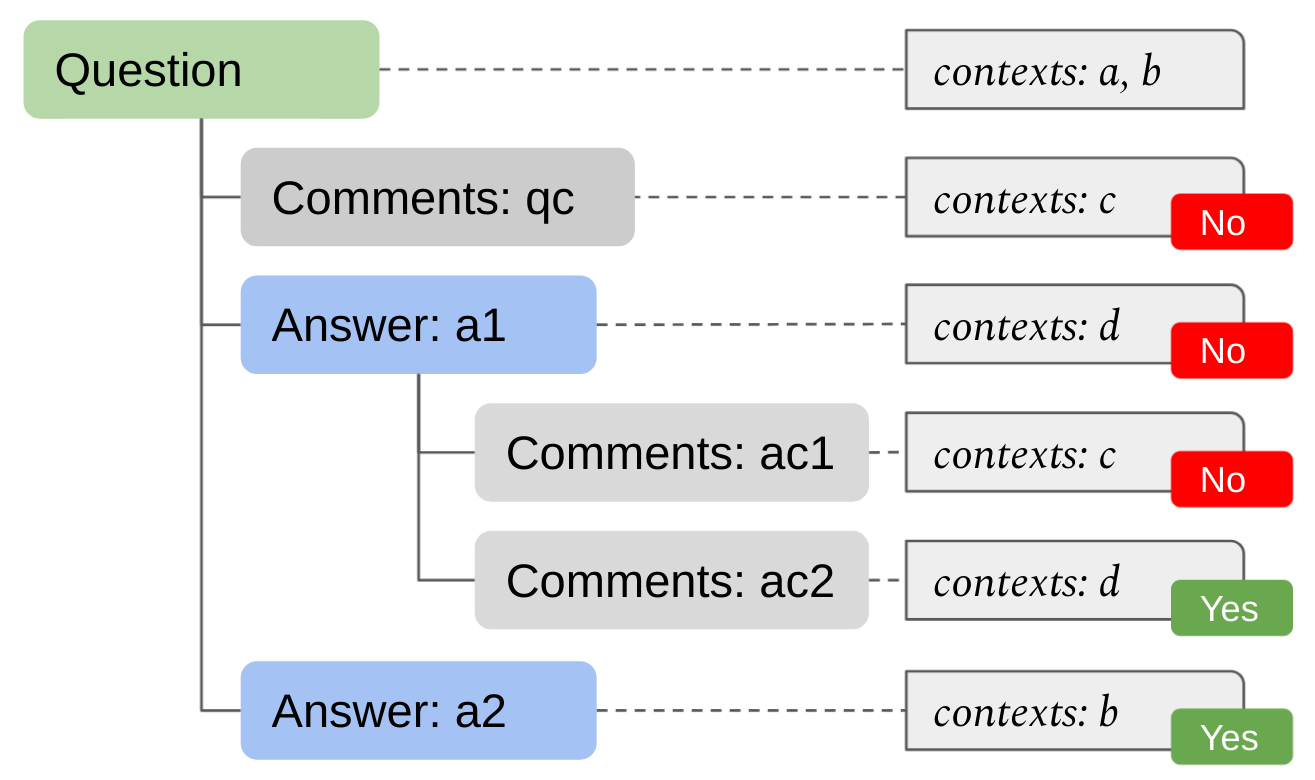}
 \caption{Stack Overflow thread hierarchy to determine tag overlap. \textit{Yes} means this context is already previously mentioned in a parent, while \textit{No} indicates no overlap.}
 \label{fig:overlap}
\end{figure}

Figure \ref{fig:overlap} shows the hierarchical structure of a thread where the question is at the top level, question comments and answers are at the next level, and answer comments are at the bottom of the hierarchy. Assume we extracted the contexts \emph{a, b, c} and \emph{d} in the various parts of the thread as shown in the Figure \ref{fig:overlap}. 
When we are reviewing comment \emph{ac2}, we would select \textit{Yes} for the overlap question since \emph{d} exists in answer \emph{a1}, which is an immediate parent of the comment.
On the other hand, when reviewing the same context \emph{d} for answer \emph{a1}, an annotator would mark \textit{No} overlap since \emph{d} was not mentioned in any of \emph{a1}'s parents.
Following this guide, in Figure~\ref{fig:context_distribution}, a reviewer would mark no overlap for sentence 1 but yes for sentence 3 since \textit{static} does appear in the code snippet posted in the question even though the tag is used as intended.
Note that the answers to Q1 and Q2 determine whether a sentence is considered as containing additional context or not.
Specifically, we consider a sentence as containing additional context if the tag is used as intended (Q1=Yes) and there is no overlap with any of its parents (Q2=No).
In Table~\ref{tab:annotation_ex}, Sentence 1 is the only sentence containing additional context.

\paragraph{Coding guide for annotating question 3 (Q3):} 
To understand why the identified tags are mentioned as additional context, we ask annotators to provide a free-form text response for Q3. Annotators read the surrounding text to understand the purpose of the tag in the sentence.  
We did not provide any guidance for answering this question, because we did not want to influence the annotators' response.
We later use this information in Phase \tc{3}.

\paragraph{Distributing threads between annotators:}
As mentioned earlier, we manually review a statistically representative sample of 69 threads from each tag, totalling 207 threads. To ensure the reliability of study, we use an iterative process for manually coding the data, starting with a general guideline for capturing additional contexts. Instead of annotating all threads at once, we conduct two pilot rounds using only a subset of threads from each tag. In the first round, three authors annotate the same 30 threads. We calculate the Fleiss' kappa score \cite{Kappa:2015} for Q1 and Q2 to determine the ambiguity of the coding task. 
We then discuss any disagreements and use the insights from our discussions to remove ambiguities in the coding guide in each round.
The three authors then use the refined coding guide to annotate another 30 threads.
Table \ref{tab:agreement} shows the statistics of the first two rounds of annotations. Note that we randomly select ten threads from each tag for the first two rounds (a total of sixty threads). However, Table \ref{tab:agreement} only counts the threads and sentences that the automated detection process from Phase \tc{1} flagged as candidates for review. For example, in round 1, only six out of ten \emph{Json} threads had potential additional contexts. 
After round 2, we were satisfied with the moderate agreement level~\cite{altman:1990} (0.416). This level of agreement is not surprising since ``additional context'' is a new concept without existing coding schema. Also, we annotated questions Q1 and Q2 at the same time in each round. Therefore, disagreement in Q1 results in disagreement in Q2.
We then distribute the remaining 147 among the three authors to review individually.

The result of Phase \tc{2} is a set of confirmed sentences with additional context, which allows us to answer RQ1.

\begin{table}[t!]
    \centering
    \caption{Fleiss kappa scores for Q1 and Q2 of first two annotation rounds. Since Q2 is answered only if Q1=\textit{Yes}, we show the number of sentences annotated in each round for each question.}
    \resizebox{0.4\textwidth}{!}{
	\begin{tabular}{llrrrrr}
		\toprule
		\multirow{2}{*}{{\textbf{Round}}} & 
		\multirow{2}{*}{{\textbf{Tag}}} & 
		\multirow{2}{*}{{\textbf{Threads}}} &
		\multicolumn{2}{c}{\textbf{Q1 annotation}}&
		\multicolumn{2}{c}{\textbf{Q2 annotation}}\\
		\cmidrule(l){4-5}\cmidrule(l){6-7}
		&&&
		\multicolumn{1}{c}{\textbf{\#}} & 
		\multicolumn{1}{c}{\textbf{Kappa}} & 
		\multicolumn{1}{c}{\textbf{\#}} & 
		\multicolumn{1}{c}{\textbf{Kappa}} \\
		\midrule
		\multirow{4}{*}{1} & Json & 6 & 35 & 0.91 & 23 & 0.137 \\
		                   & Django & 8 & 46 & 0.613 & 30 & 0.273 \\
		                   & Regex & 10 & 96 & 0.809 & 64 & 0.106 \\
		                   \cmidrule{2-7}
		                   & Overall & 24 & 177 & 0.783 & 117 & 0.223 \\
		\midrule
		\multirow{4}{*}{2} & Json & 7 & 73 & 0.69 & 39 & 0.319 \\
		                   & Django & 5 & 17 & 0.664 & 8 & 0.7 \\
		                   & Regex & 6 & 37 & 0.881 & 11 & 0.542 \\
		                   \cmidrule{2-7}
		                   & Overall & 18 & 127 & 0.757 & 58 & 0.416 \\
		\bottomrule	
	\end{tabular}
	}
	\label{tab:agreement}
	
\end{table}

\subsection{Phase \tc{3} Qualitative Analysis of Additional Context}
\label{sec:phase3}
In Phase \tc{3}, we answer RQ2 and RQ3. Once we determine that a tag mentioned in a sentence is an additional context in Phase \tc{2}, we analyze the tags to determine their high-level technological category (RQ2) and why the additional context is mentioned (RQ3).

\subsubsection{Identifying categories of additional context}
To answer RQ2, we conduct open card sorting \cite{OpenCoding:2016} of the additional contexts identified in Phase \tc{2}. This technique allows us to iteratively arrange cards (tags) into groups so that we can understand the types of additional context. We create cards for all tags that are identified as additional contexts. 
Three authors then work together to group tags that have similar characteristics, based on their tag descriptions.
They then provide a descriptive name for each category.
For example, \emph{java}, \emph{r}, and \emph{c++} are categorized as programming language. 
Note that we did not rely on tag categories in \emph{Witt} for this task, because there is more than one possible category for each tag. 
For example, in Witt, the tag \emph{django} is categorized as ``software'', ``framework'', and ``web-framework''.

\subsubsection{Identifying reasons for mentioning additional context}
To determine categories of reasons for mentioning additional context, we use the annotators' free-text answers to Q3 from Phase \tc{2}. 
We collect all free-form responses of Q3 from all annotation rounds and the three authors perform an open coding card sorting of these reasons.
Once we determine categories of reasons, we ask each author to answer Q3 again, but this time using closed coding where they have to select one of the pre-determined categories.
To ensure our categories and their definitions are clear for the closed-coding analysis, we first conduct a pilot round using only a subset of threads from our sample set of threads and using three authors for each sentence.
Then we refine the coding guide by providing examples of when to use each code, because we observed that annotator agreement is low during the pilot round. After refining the coding guide, we assign two authors to annotate each of the remaining sentences. Disagreements in round 2 were handled through discussions involving all three authors. Table \ref{tab:res_agreement} shows the agreement rate between raters in each round.

\begin{table}[t!]
    \centering
    \caption{Kappa scores for reasons for using ``tag/context'' in a sentence (Q3, Phase \tc{3}). We use Fleiss Kappa Score for annotations in both rounds.}
	\begin{tabular}{llrr}
		\toprule
		\multicolumn{1}{c}{\textbf{Round}} & 
		\multicolumn{1}{c}{\textbf{Raters}} &
		\multicolumn{1}{c}{\textbf{\# Sentences}} & 
		\multicolumn{1}{c}{\textbf{Kappa Score}} \\
		\midrule
		\multirow{1}{*}{1} & 1-2-3 & 105 & 0.354  \\
		\midrule
		\multirow{3}{*}{2} & 1-2 & 144 &  0.83  \\
		                   & 2-3 & 146 &  0.847  \\
		                   & 1-3 & 144 &  0.777  \\
		\bottomrule	
	\end{tabular}
	\label{tab:res_agreement}
	
\end{table}

\section{Empirical Study Results}
 
\subsection{RQ1: Frequency of additional context in \SO{} threads}

\begin{figure}[t!]
 \centering
 \includegraphics[scale=0.48]{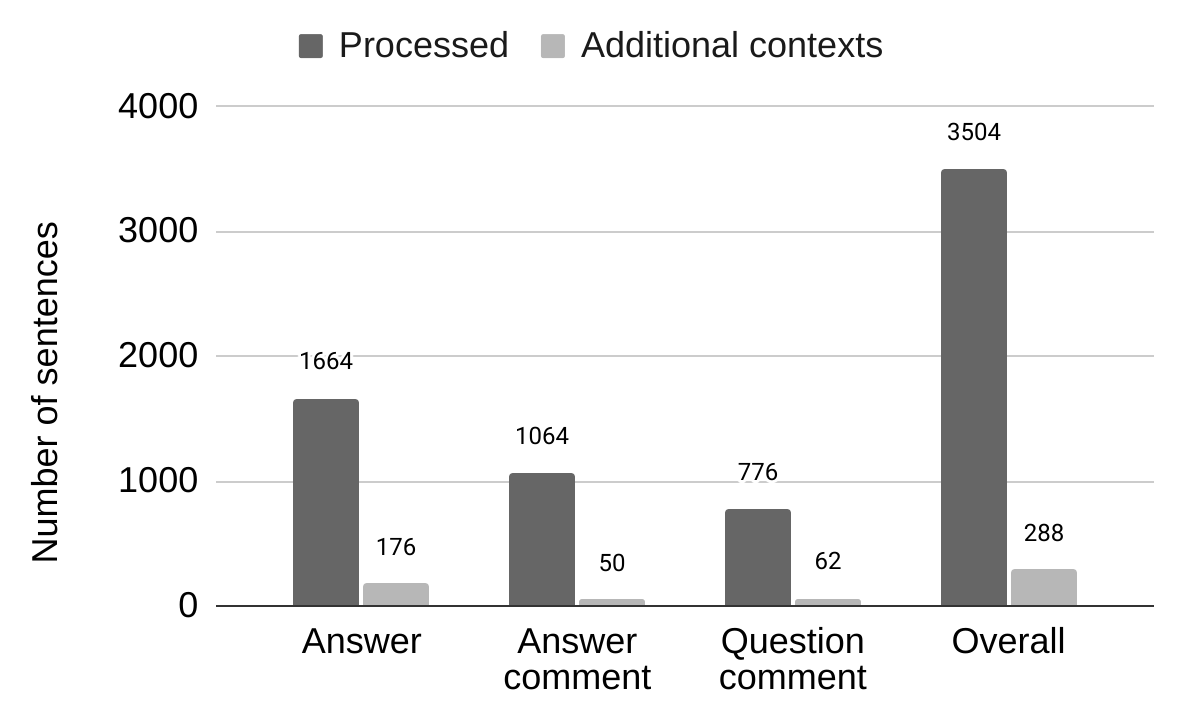}
 \caption{Comparison of sentences with additional contexts to total number of sentences processed.}
 \label{fig:sentence_proportions}
\end{figure}

Overall, we analyze 207 threads in our empirical study, containing a total of 3,504 sentences.
The ``Processed'' (dark grey) bars in Figure~\ref{fig:sentence_proportions} show the distribution of these processed sentences across answers, answer comments, and question comments.
Out of these 3,504 sentences, the automatic identification process of Phase \tc{1} flagged 595 sentences for review.
We then manually reviewed these 595 sentences in Phase \tc{2}.
To answer RQ1, we count the sentences that were annotated with \textit{Yes} to Q1 and \textit{No} to Q2 in Phase \tc{2}. 

\begin{figure}[t!]
 \centering
 \includegraphics[scale=0.38]{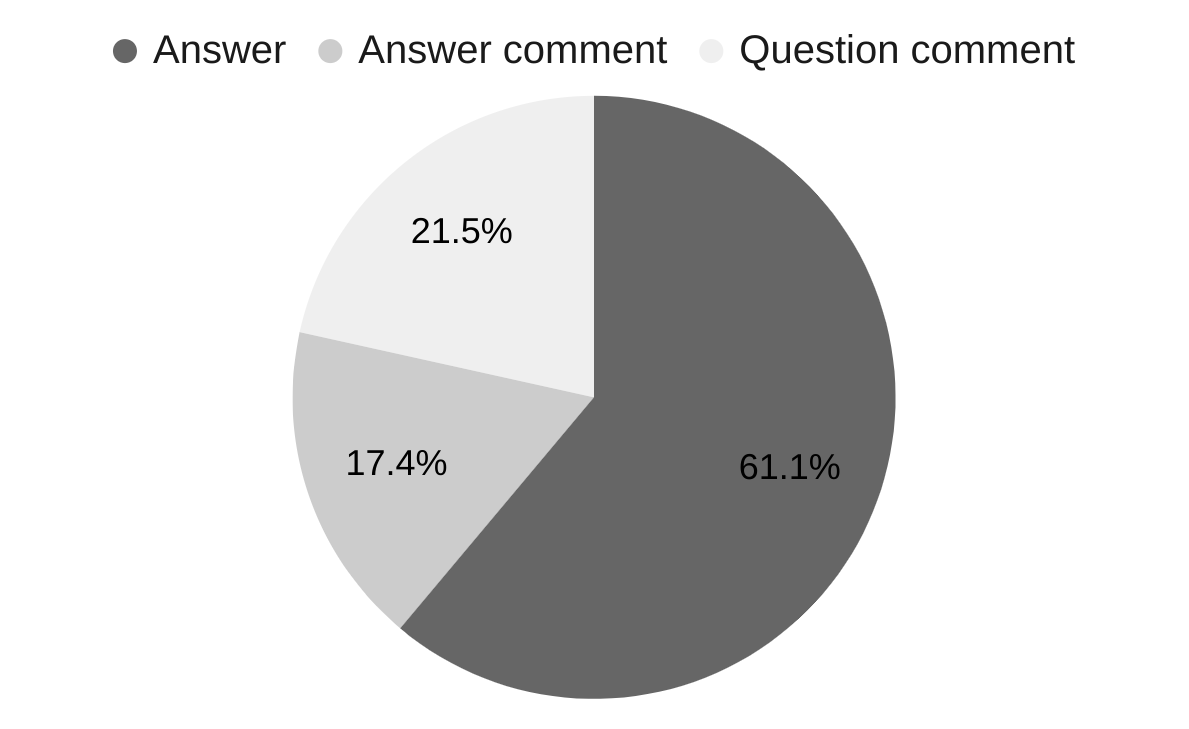}
 \caption{Location of the 288 confirmed sentences with additional context.}
 \label{fig:context_distribution}
\end{figure}

Figure \ref{fig:sentence_proportions} shows how many sentences contain additional context (light grey bars) compared to the total number of processed sentences. Overall, out of the 3,504 processed sentences, only 288 ($\sim$8\%) sentences contain additional contexts. 
As shown in Figure~\ref{fig:context_distribution}, the majority of the 288 sentences with additional context were found in answers (61.1\%). 
The remaining sentences with additional context appear in question comments (21.5\%) and answer comments (17.4\%), respectively.
Note that these 288 confirmed sentences belong to 96 unique threads. This means $\sim$46\% of the 207 threads that we study contain at least one sentence with additional context.

\begin{findingenv}{RQ1 Summary}{finding:rq1}
While only $\sim$8\% of the total sentences processed contain additional context, 46\% of the threads we analyze contain at least one sentence with additional context. 
The majority of sentences with additional context (61.1\%) occur in answers.
\end{findingenv}

\subsection{RQ2: Types of additional context in \SO{} threads}

The 288 confirmed sentences with additional contexts in RQ1 contain 142 unique tags.
Through open card sorting, we create 16 technical categories to describe these tags. Table \ref{tab:tag_categories} shows the categories, the number/proportion of tags that belong to each category, and their distribution in sentences with additional contexts. The complete list can be found in our artifact page\footnote{https://figshare.com/s/97007e09150ca97046a6}.

The top five categories in Table \ref{tab:tag_categories} account for $\sim$79\% of the tags we found as additional contexts. The rest of the tags are spread across the remaining eleven categories. Nearly 38\% of the additional contexts that we find are either a ``computing concept'', such as \emph{command}, or a ``programming concept'', such as \emph{package}. The third column of Table \ref{tab:tag_categories} shows the distribution of tag categories in sentences with additional contexts. Computing concepts and programming concepts account for approximately 46\% of the sentences with additional contexts while the other top four categories account for 45\% of the sentences.
Apart from concepts, the most common types of additional contexts are programming languages (17\% of sentences) and library/frameworks (11\% of sentences).
Overall, the categories library/framework, programming language, tool/application, API and database cover 50\% of the additional contexts and are found in 45\% of the sentences with additional contexts.

\begin{findingenv}{RQ2 Summary}{finding:rq2}
38\% of the tags used as additional context are either computing concepts or programming concepts. Categories such as library/framework, programming language, tool/application, API and database cover 50\% of the types of additional contexts and are found in 45\% of the sentences with additional contexts. 
\end{findingenv} 

\begin{table}[t!]
    \centering
    \caption{The 16 categories of tags that appear in the 288 sentences with additional context. We show the number (proportion) of tags in each category, as well as the number (proportion) of sentences.}
    
    \resizebox{0.4\textwidth}{!}{
	\begin{tabular}{lrrl}
		\toprule
		 \multicolumn{1}{l}{\textbf{Tag category}} & 
		 \multicolumn{1}{c}{\textbf{Tag }} &
		 \multicolumn{1}{c}{\textbf{Sentence }} &
		 \multicolumn{1}{l}{\textbf{Example}} \\
		 &
		 \multicolumn{1}{c}{\textbf{count (\%)}} &
		 \multicolumn{1}{c}{\textbf{Count (\%)}} & \\
		\midrule
		Computing concept & 32 (23\%) & 96 (31\%) & command\\
		Library/framework & 26 (18\%) & 35 (11\%) & asp.net\\
		Programming concept & 22 (15\%) & 47 (15\%) & package \\
		Programming language & 16 (11\%) & 52 (17\%) & perl  \\
		Tool/application & 16 (11\%) & 26 (8\%)  & browser \\
		API & 8 (6\%) & 17 (5\%) & tostring \\
		Database & 5 (4\%) & 11 (4\%) & oracle \\
		Standard & 3 (2\%) & 6 (2\%) & wsgi \\
		Protocol & 3 (2\%) & 9 (3\%)  & http  \\
		Data format & 2 (1\%) & 3 (1\%) & cbor \\
		IDE & 2 (1\%) & 2 (1\%) & pyscripter \\
		Web server & 2 (1\%) & 4 (1\%) & iis \\
		Template engine & 2 (1\%) & 2 (1\%) & razor \\
		Technology stack & 1 (1\%) & 1 (0\%) & xampp \\
		Project & 1 (1\%) & 1 (0\%) & gnu \\
		Hosting service & 1 (1\%)  & 2 (1\%) & github \\
		\bottomrule	
	\end{tabular}
	}
	\label{tab:tag_categories}
	
\end{table}

\subsection{RQ3: Purpose of additional contexts}
Our open-coding of the free-text reasons provided for A3 resulted in ten reasons as to why additional contexts are mentioned in a give thread.
Table \ref{tab:tag_reasons} shows the distribution of reasons across the 288 sentences. We first explain each category, along with examples from \SO{} threads.


We find that in many cases, the \cont{} is mentioned as a direct solution to the posted problem. For example, \textit{Solution - suggest API/library/framework} includes sentences where a user suggests an API/library/framework as a direct solution to a problem, e.g., ``\textit{Use a StringBuilder instead}'' from \qid{25803443} contains the tag \emph{stringbuilder}. This sentence directly suggests  the \verb|StringBuilder| API as a solution to the posed problem.
Similarly, in the 
\textit{Solution - suggest programming language} category, the purpose of the \cont{} is to suggest a programming language to directly solve the problem. The tag \emph{javascript} used in ``\textit{Or go full javascript and add <p> element...}'' from \qid{24261644} is an example of this case.
We use the category \textit{{Solution - own experience}} when the technical context is a direct solution that is simply based on personal experience, not necessarily a unique solution. For example, \emph{PyScripter} in ``\textit{PyScripter works for me ..}'' from \qid{11699407} states a solution based on a personal experience.
Finally, we use \textit{Solution - other} for any other \conts{} that is mentioned as a direct solution but that does not fall into any of the above solution categories. For example, \qid{26981520} contains the sentence ``\textit{As for the type of DBMS, a NoSQL would be recommended, like MongoDB}'' which has the additional context \emph{mongodb} that provides a direct solution but that does not fit the above categories.

We also find that several of the \conts{} are mentioned as solution conditions, i.e., the suggested solution works only under specific condition. Note that these categories are different from the \textit{Solution} categories as they are not direct solutions.
The first of these categories is 
\textit{Solution condition - programming language} where the solution holds only for a specific programming language. \emph{javascript} in ``\textit{Example in JavaScript:}'' from \qid{33783429} is an example of a sentence in this category.
The category \textit{Solution condition - version} contains sentences where the tag is mentioned to denote a specific version (typically a programming language or library/framework) that is a condition for the answer. For example, the tag \emph{django 1.6} in \textit{``For django 1.6, in settings...''} (\qid{2551933}) suggests that this answer applies only to this specific version of django.
We use the category 
\textit{Solution condition - environment} when the tag is used to denote the expected environment for this solution to work, such as operating system, browser, or technology stack. For example, ``\textit{In case you are dealing with a legacy system...}'' (\qid{1812891}) states a solution that only holds for old systems.
Finally, 
\textit{Solution condition - other} encompasses other solution conditions that do not fall into the above cases. For example, \emph{postgresql} in \textit{``If you use a PostgreSQL database, you can do this simpler ...''} (\qid{56201268}) is a solution condition; using PostgreSQL as the database is not essential to answering this question, but the given solution in this answer only works with PostgreSQL. We argue that \textit{solution condition} is the most interesting type of additional context since it indicates that a solution only applies in a certain situation. While it is possible to merge some of the subcategories of \textit{solution condition} (or \textit{solution}), the purpose of this analysis is to identify the intent of the tag mentioned in sentences.

When analyzing the reasons for mentioning \conts{} in sentences, we find that there are many cases that mention the tag as part of a technical explanation that describes a solution or elaborates details, whether directly or through external references. 
We also find cases where the tag represents essential or trivial terms that are necessary to the discussion of the thread (e.g., ``\textit{valid JSON format requires keys to be strings in double quotes}'').
We group all these cases under one category 
 \textit{Technical explanation/discussion/resource}.
Finally, we include an \textit{Other} category for sentences with additional contexts that did not match with any of the previous categories.

We find that most of the additional context in the 288 sentences is mentioned as part of a technical explanation/discussion or as an additional resource.
This reason alone covers 79\% of the 288 sentences.
We find that 17\% of the time, the additional context is a direct solution to the thread topic, including an API/library/framework or programming language.
Such additional context would be valuable for developers navigating the thread.
In 8\% of the sentences, the additional context provides specific conditions that need to be satisfied for this solution to be useful to a developer.
These include specific APIs or configuration environments, which developers are searching for~\cite{DevHelp:2013}.


We also compare the reason a specific tag is mentioned and the category of this tag from Table~\ref{tab:tag_categories} from RQ2.
We find that 94\% of sentences whose tags belong to the programming concepts or computing concepts categories are categorized as technical explanation/discussion/resource in Table~\ref{tab:tag_reasons}. 
We also find that 40\% of the tags in the library/framework, programming language, tool/application, API, database, and web server categories in Table~\ref{tab:tag_categories} are in one of the solution categories in Table~\ref{tab:tag_reasons}.
This suggests that the tag category may often help in determining the purpose of mentioning a specific tag in a sentence, which in turn can help determine if it can serve as a navigational cue.
We discuss these implications more in Section~\ref{sec:discussion}.

\begin{table}[t!]
    \centering
    \caption{Distribution of reason categories in sentences with additional contexts.}
	\begin{tabular}{lr}
		\toprule
		 \multicolumn{1}{l}{\textbf{Reason category}} & 
		 \multicolumn{1}{c}{\textbf{Sentence }}  \\
		 &
		 \multicolumn{1}{c}{\textbf{count (\%)}}  \\
		\midrule
		Technical explanation/discussion/resource & 224 (79\%)  \\
		Solution - suggest API/library/framework & 16 (6\%)  \\
		Solution condition - version & 13 (5\%)  \\
		Solution - own experience & 8 (3\%)  \\
		Solution condition - environment & 7 (2\%)  \\
		Solution - other & 7 (2\%)  \\
		Other & 4 (1\%)  \\
		Solution condition - programming language & 3 (1\%)  \\
		Solution - suggest programming language & 2 (1\%) \\
		Solution condition - other	& 1	(0\%)  \\
		\bottomrule	
	\end{tabular}
	\label{tab:tag_reasons}
	
\end{table}

\begin{findingenv}{RQ3 Summary}{finding:rq3}
79\% of additional contexts are used to indicate necessary technical explanation or resource. The remaining approximately 20\% are direct solutions or solution conditions.
\end{findingenv}

\section{Discussion}
\label{sec:discussion}

The goal of our empirical study is to understand how often additional context occurs in Stack Overflow answers/comments, as well as the types of additional context and their purpose in a thread.
The results of such a study provide insights into whether additional context, i.e., technical information mentioned elsewhere in a thread and not already in the question, can potentially be used as navigational cues.

\begin{figure}[t!]
 \centering
 \includegraphics[scale=0.6]{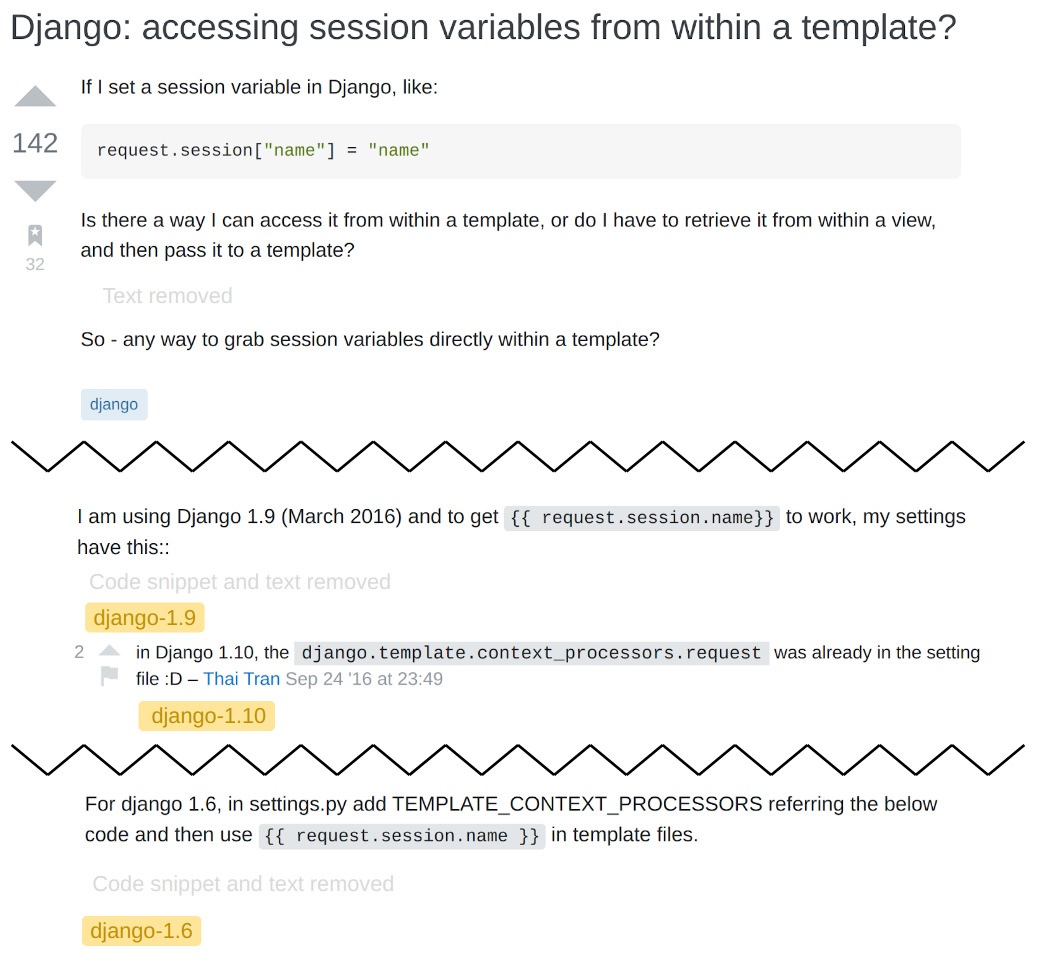}
 \caption{Emphasizing additional context through answer/comment tags that can serve as navigational cues
 }
 \label{fig:implication}
\end{figure}

\paragraph{Implications for Navigation} Our findings show that approximately half of the threads we analyzed contain at least one sentence with additional context. 
This means that in almost half of the threads, there is potential for providing additional navigational cues.
Additionally, those sentences account for only $\sim$8\% of all sentences in these threads.
This is a positive indication for navigational cues; if a large proportion of sentences contained additional context, using all of them as cues could clutter the thread instead of helping users identify the information they are looking for. 
Overall, we envision that one way for adding navigational cues is by representing the additional context we studied as answer/comment tags as shown in Figure~\ref{fig:implication}. Once a user finds a relevant question/thread, they can use these answer/comment tags to manually navigate to relevant information on the thread (i.e. quickly scrolling to a relevant answer). That is, we do not intend to make use of answer/comments tags for searching threads. Currently, \SO{} uses question tags to help users search for relevant threads. Therefore, the added answer/comment tags should not divert the search engine as they are not part of the question tags.

\paragraph{Implications for Automated Creation of Navigational Cues}
Ideally, the navigational cue tags we envision in Figure~\ref{fig:implication} would be automatically captured and created without the need for their authors to add them.
This allows the addition of navigational cues to existing \SO{} content.
In our work, we used tags as proxies for identifying additional context.
Based on the findings of RQ2 and RQ3, there are clearly certain types of tags that are more likely to serve as useful navigational cues than others.
Specifically, tags in the programming concept and computing concept categories are typically mentioned as part of a general discussion/explanation that does not warrant tagging the whole answer with that tag.
On the other hand, tags in categories such as API, library/framework, programming language, or web server were more likely to be about a direct solution or condition that is necessary for this solution to work.
A developer navigating a thread might already be looking to solutions related to a specific technology or those that match their technology stack and requirements.
Thus, creating navigational cues for such tag categories is more promising.
Future research into automatically, and precisely, categorizing tags into the categories we identified can be helpful for adding navigation cues.
Such automatic categorization along with our automated identification of additional context (Phase \tc{1}) can then be used to augment \SO{} threads with add navigational cues.

\section{Threats to Validity}

\paragraph{Internal validity:} We automatically capture tags mentioned in sentences to reduce the manual review burden on annotators. Instead of using the complete tag set from \SO, we use a subset of tags extracted from the \emph{Witt} taxonomy \cite{Witt:2019}. Since the \emph{Witt} taxonomy was created in 2019, our automated approach could miss newer tags mentioned in sentences that are not available in \emph{Witt}. However, the goal of our study is to understand the usage of technical context and the general phenomena of additional context, not to automatically identify them. Therefore, using \emph{Witt} does not affect our understanding of additional context; our results generally represent a lower bound of the occurrence of additional context.

We use the Stanford CoreNLP toolkit~\cite{CoreNLP:2014} to extract sentences from answers and comments. Any limitations of the tools affect the sentence extraction step of Phase \tc{1}. 
For example, the CoreNLP sentence processor may not properly parse poorly written or poorly punctuated sentences, leading to some of the reported sentences not being complete sentences.
Since our goal is to understand the occurrence of additional context, searching for it in sentences with grammatical mistakes does not affect our results, especially since we manually validate all data.


When identifying additional context, three authors first annotate $\sim$29\% of the threads from the sample. The rest of the threads were divided among each author to annotate individually, which is standard practice when annotating large data sets~\cite{SOGoodCode:2012, Annotate2:2019, Annotate1:2021}.
Doing the initial round of annotation with three authors ensures common understanding of the coding guidelines.
We also calculate and report inter-rater agreement for two of the annotation questions (Q1 and Q2). In each pilot annotation round, we observed substantial (>0.61)~\cite{altman:1990} agreement for Q1. Even though we observed a fair (0.223) score for Q2 in round 1, there was an improvement in agreement score in round 2 (0.416) once we removed the ambiguity in the annotation guide. 
In the pilot round, when all three authors annotate Q3, our inter-rater agreement for close coding was moderate (0.354)~\cite{altman:1990}. After refining the coding guide, the remaining sentences were divided among three authors such that each sentence is coded by two authors. Our inter-rater agreement between each author was substantial (>0.7)~\cite{altman:1990}.
The systematic process we followed and the increase in agreement scores gives us confidence in the reliability of the individually annotated data.
\paragraph{Construct validity:} We use \SO{} tags as a proxy for technical context. Since \SO{} tags are user created contents, there could be multiple token-level representation for the same tag (e.g. \emph{JS} and \emph{Javascript}). The intention of automatically capturing tag mentions is to reduce the manual review burden of annotators from reading all sentences. 
While this proxy could exclude some of the sentences that contain additional contexts from being manually examined, our goal is to conduct an initial empirical study to understand the new concept of additional context before investing in efforts that capture all possible representations of technical contexts.

Some \SO{} tags may be used with a different intention in the sentences. This is why we did not rely solely on the automatic identification of technical context but chose to manually confirm all results. However, there may also be other important contextual information that can be used for navigation that goes beyond technical context (e.g., non-functional properties such as performance or security being mentioned in threads). We choose to limit the scope of this study to technical context to avoid the subjectivity of deciding what other contextual information may be~\cite{Nadi:2020}.
Future work that includes human participants is needed to conclude that the additional navigational cues based on technical context will help developers find information faster; we do not claim to measure this in our work.
Instead, we study how additional context is used to understand its potential to serve as navigational cues.

Some of the accepted answers in our data did not contain additional contexts, which can lead to the argument that adding visual cues for additional context can mislead users to unverified answers. However, answers other than the accepted one are not necessarily incorrect. Similar to the checkmark indicating that an answer is the one selected by the original poster (not necessarily the only correct answer \cite{Zhang:2021}), an additional context visual cue will alert users that an answer/comment is specific to a particular context (e.g., an OS) and allow users to dismiss irrelevant content (e.g., about other OSs).

\paragraph{External validity:} When collecting threads for annotation, we limit our study to threads from three tags: \verb|json|, \verb|regex|, and \verb|django|. Even though we focused on threads from only these tags, the threads in our sample contain 142 unique technical contexts from different programming languages, databases, libraries/frameworks, and APIs. We do not claim that our study encompasses \textit{all} possible additional context, but it provides first insights into the occurrence and usage of additional context in \SO{} threads.

\section{Conclusion}

Given the vast amount of knowledge available on \SO, finding the right information that applies to developers' specific requirements or technology stack is challenging. In this paper, we define \textit{context} as information that characterizes the technologies and assumptions of a given question or answer.
We define \textit{additional context} as context in a thread's answer or comment that does not overlap with the question context.
We argue that such additional context may provide useful navigational cues for \SO{} users.
In this paper, we conduct an empirical study of the occurrence and nature of additional context on \SO{} to understand its potential for being used as navigational cues.
We find that the majority of the additional contexts are computing concepts that discuss necessary technical details of the thread. However, technology categories such as library/framework, programming language, API, and tool/application usually provide a solution to the original problem. 
Since only a small percentage of threads contain additional context that provide a solution, we could leverage these findings to automatically identify such technical contexts to generate navigational cues. 


\bibliographystyle{ACM-Reference-Format}
\bibliography{references}



\end{document}